\documentclass[aps,prl,reprint,superscriptaddress]{revtex4-1}
\usepackage{graphicx}
\usepackage{rotating}
\usepackage{amssymb}    
\usepackage{amsmath}   
\usepackage{epsfig}
\usepackage[normalem]{ulem}   
\usepackage{bm}   
\usepackage{color}
\newcommand{\e}{\mathrm{e}}

\date{\today}

\begin{document}

\title{Torsion-Adding and Asymptotic Winding Number for Periodic Window Sequences}

\author{E. S. Medeiros}
\email{esm@if.usp.br}
\affiliation{Instituto de F\'isica, Universidade de S\~ao Paulo, S\~ao Paulo, Brazil}
\author{R. O. Medrano-T}
\affiliation{Departamento de Ci\^encias Exatas e da Terra, Universidade Federal de S\~ao Paulo,  S\~ao Paulo, Brazil}
\author{I. L. Caldas}
\affiliation{Instituto de F\'isica, Universidade de S\~ao Paulo, S\~ao Paulo, Brazil}
\author{S. L. T. de Souza}
\affiliation{Departamento de F\'isica e Matem\'atica, Universidade Federal de S\~ao Jo\~ao del-Rei, Minas Gerais, Brazil}

\begin{abstract}

In parameter space of nonlinear dynamical systems, windows of periodic states are aligned following routes of period-adding configuring periodic window sequences. In state space of driven nonlinear oscillators, we determine the torsion associated with the periodic states and identify regions of uniform torsion in the window sequences. Moreover, we find that the measured of torsion differs by a constant between successive windows in periodic window sequences. We call this phenomenon as torsion-adding. Finally, combining the torsion and the period adding rules, we deduce a general rule to obtain the asymptotic winding number in the accumulation limit of such periodic window sequences.

\end{abstract}


\maketitle

A conspicuous characteristic in parameter space of dissipative nonlinear dynamical systems is the appearance of periodic states for parameter sets immersed in parameter regions correspondent to chaotic states. In the literature, much attention has been devoted to establish connections between these periodic states. For example, a successive constant increment on the period of oscillation of such states (period-adding phenomenon) \cite{Kaneko1982} have been experimental and numerically observed in several real-world systems such as neuronal activities \cite{Ren1997,Jia2012}, electronic circuits \cite{Periodchua1986}, bubble formation \cite{bubblesarto2004}, semiconductor device \cite{Yasuda1994}, and chemical reaction \cite{Hauser1997}. The period-adding phenomenon has been also observed for sequences of shrimp-shaped periodic windows accumulating in specific parameter space regions \cite{Bonatto2007, Albuquerque2012,Stegemann2010, Gallas1993, Murilo1996, Stoop2010}. Once nonlinear dynamical systems can exhibit many different kinds of motion, knowing adding rules, such as the period-adding and further information about the accumulating parameter regions, is very advantageous, specially, for predicting periodic states for different parameter sets in real-world applications.

Furthermore, besides the intrinsic period of oscillations, in dissipative systems, periodic states have other interesting convergence properties. For instance, for driven nonlinear oscillators, the {\it torsion number} $n$ is defined as number of twists that local flow perform around a given periodic solution during a dynamical period $m$, and { \it the winding number} defined as $w=n/m$ \cite{Parlitz1985a, Parlitz1985b, Parlitz1990, Englisch1991, Parlitz1993}. However, besides the existence of such convergence properties, additional connecting rules between periodic states and accumulating regions characteristic have not yet been discovered.

Our aim here is to investigate the convergence characteristics, namely, the torsion and winding number of periodic states within complex periodic windows, in period-adding sequences in the parameter space of driven nonlinear oscillators. A torsion-adding formulation between such periodic states are proposed here. Combining both additive sequences properties, the torsion and the period adding, we describe a generic periodic window in a sequence in terms of its winding number. The asymptotic limit of such description gives a general rule to determine the winding number for any accumulation of period-adding sequences.

\begin{figure}[!htp] \centering
\includegraphics[width=8.5cm,height=4cm]{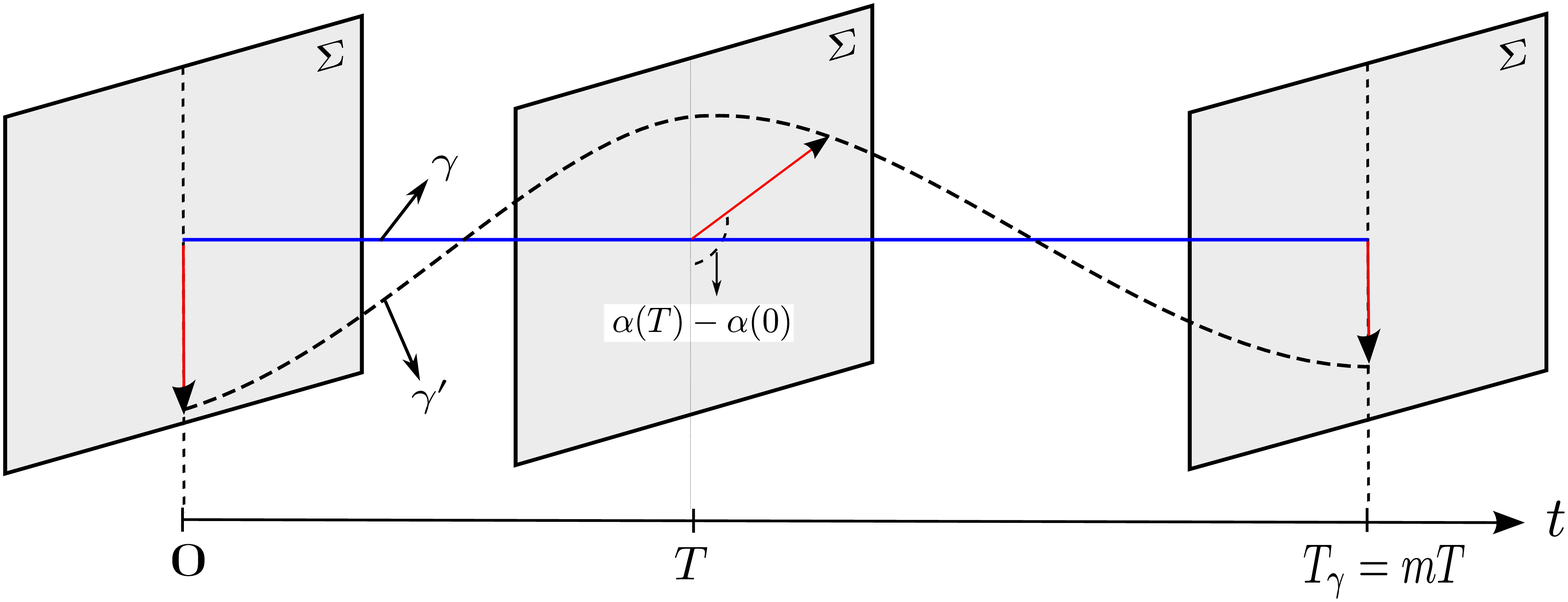}
\caption{(Color online) Sketch of a periodic orbit $\gamma$ with period $T_\gamma$ surrounded by a neighbor orbit $\gamma'$, $\alpha(t)$ is the phase of the $\gamma'$ revolution around $\gamma$. The stroboscopic Poincar\'e section $\Sigma$ is defined by the forcing period $T$.}
\label{fig:1}
\end{figure}

Generally, the driven nonlinear oscillator is described by:

\begin{equation}
  \ddot{x}+g(x,\dot{x}) = h(t),
  \label{eq:1}
\end{equation}
where $h(t)=h(t+T)$ is a periodic function with angular frequency $\omega=2\pi/T$. For this equation, the winding number is obtained by considering revolutions performed by an orbit $\gamma'$ around of a very close neighbor periodic orbit $\gamma$ during the interval time $\Delta t = T$ (see Fig. \ref{fig:1}). The absolute mean value of the revolution angular frequency, $\Omega(\gamma) = |<\dot{\alpha}(t)>|$, is called {\it torsion frequency}:
\begin{equation} 
  \Omega(\gamma) =
  \lim_{t \rightarrow  \infty} {\frac{1}{t} \left|\int_0^t \!
    \dot{\alpha}(t') \, \mathrm{d}t'\right|} = 
  \lim_{t \rightarrow \infty} {\frac{|\alpha (t) - \alpha(0)|}{t}}.
  \label{eq:2}
\end{equation}
Thus, considering the $T$-shift map \cite{Nota1}, the winding number is precisely defined as \cite{Parlitz1985a}:
\begin{equation}
  w(\gamma) = \frac{\Omega (\gamma)}{\omega}.
  \label{eq:3}
\end{equation}
For a more appropriated form, note that the $\gamma$ period is given by $T_{\gamma} = mT = 2\pi m/\omega$, while the torsion frequency period is $T_{\Omega} = 2\pi/\Omega$ (see Fig. \ref{fig:1}). Including these periods in Eq. (\ref{eq:3}), and defining the {\it torsion number} as $n = T_{\gamma}/T_{\Omega}$, we obtain the winding number
\begin{equation}
  w(\gamma) = \frac{n}{m}.
  \label{eq:4}
\end{equation}

Figure 1 here

\begin{figure}[!htp] \centering
\includegraphics[scale=0.25]{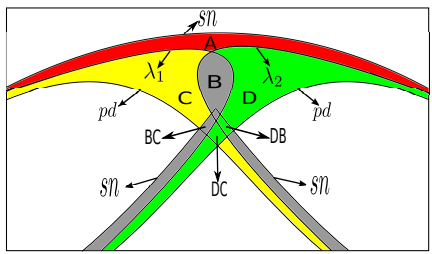}
\caption{(Color online) Main-body window in two-dimensional parameter
  space. $sn$ and $pd$ are the saddle-node and period-doubling
  bifurcation curves. The curves $\lambda_1$ and $\lambda_2$ are the
  skeleton of the structure delimiting the regions $A$ (red), $B$ (gray), $C$ (yellow), and $D$ (green) are regions delimited by the skeleton.}
\label{fig:2}
\end{figure}

In Fig. \ref{fig:2}, we show a sketch of the region with the lowest period inside a complex periodic window. We refer to it as {\it main-body window}, frequently found in the two-dimensional parameter space of the dynamical system given by Eq. \ref{eq:1}. This main-body window has the same period and is bounded by saddle-node ($sn$) and period-double ($pd$) bifurcation curves. On the other hand, we identify that the winding number (Eq. \ref{eq:3}) is the same in four different regions $A$, $B$, $C$, and $D$. The point where the two saddle-node bifurcation curves met in region $B$ depicts a cusp bifurcation, where two periodic orbits are created. Bellow this point, there are limited regions where $B$, $C$, and $D$ appear in pairs ($BC$, $DB$, and $DC$), due to the coexistence of the two periodic orbits. The curves $\lambda_1$ and $\lambda_2$ between the colored regions, reveal the window {\it skeleton} \cite{Glass1984, Belair1985}. The flow converges monotonically or non monotonically to the periodic orbit according the skeleton composition ($\lambda_1$, $\lambda_2$). Crossing a $\lambda$ curve from one side to the other, the flow convergence suffers a transition. The direction of this transition characterizes $\lambda_1$ and $\lambda_2$ and causes a change in the winding number \cite{Kuznetsov2004}.

Figure 2 here

Then, we consider the winding number concept for sequences of periodic windows. Until now these sequences have been described only by period-adding rules, i.e., the period of a periodic window in the sequence can be determined by adding a constant value $\rho$ to the period of the previous window \cite{Bonatto2008a, Bonatto2008b, Souza2012, Feng2012}. Now, we introduce the torsion-adding phenomenon in periodic window sequences. In other words, for each increment $\rho$, in the window period $m_1$, the torsion number of equivalent regions is also incremented by a constant value $\tau$. Therefore, from Eq. \ref{eq:4}, the winding number $w_{R_i}$ in a generalized main-body window region $R$ ($R=A,B,C,$ or $D$) of the $i$-th window sequence can be determined in function of the torsion number in region $R$ of a known window, in particular, in function of the first one, $n_{R_1}$:
\begin{equation}  
  w_{R_i}   =  \frac{n_{R_1}+(i-1)\tau}{m_1+(i-1)\rho}.
  \label{eq:5}
\end{equation}
Thus, the asymptotic winding number limit of any region $R$,
\begin{equation}
  \lim_{i \rightarrow \infty} {w_{R_i}} = \frac{\tau}{\rho},
\label{eq:6}
\end{equation}
shows that the winding number of all regions converge to a
constant which only depends on the torsion number and the period
increments, $\tau$ and $\rho$, respectively. We denote this limit as $w_\infty$.

Another consequence of the torsion-adding is that the skeleton of all  windows in a sequence are equivalent. In fact, one can show that for $\lambda_1$ curve, in successive windows $i$ and $i+1$, the torsion-adding condition ($n_{R_{i+1}} = n_{R_i} + \tau$) imply that the torsion number difference between regions $A$ and $C$ is the same ($n_{A_i}-n_{C_i} = n_{A_{i+1}}-n_{C_{i+1}}$). Thus, the $\lambda_1$ curve promotes the same transition in all sequence.
The same is for $\lambda_2$.

Now, to verify our results, we present numerical simulations for a specific driven nonlinear oscillator, namely, the Morse oscillator that describes a diatomic molecule, immersed in an external electromagnetic field, modeled by the following differential equation \cite{Scheffczyk1991}:
\begin{equation} 
  \ddot{x}+d\dot{x}+8\e^{-x}(1-\e^{-x}) = 2.5\cos(\omega t),
\label{eq:7}
\end{equation}
where the parameter $d$ is the amplitude of the system damping and $\omega$ is the angular frequency of the external forcing. Our analysis are in the two parameter space $d \times \omega$.

As we are interested in how a trajectory $\gamma'$ converges to the stable periodic orbits $\gamma$, we consider in our numerical simulation $\gamma'$ starting in a position very close to $\gamma$ in a such way that the linearized flow
is enough to describe $\gamma'$ dynamics. We represent a point in $\gamma$ as $(x_1^*, x_2^*, x_3^*)$ and in $\gamma'$ as $(y_1, y_2, y_3)$. Therefore, the $\gamma'$ evolution is given by 
\begin{equation}
  \begin{split}
    \dot{y}_1& = y_2\\
    \dot{y}_2& = 8\e^{-x_1^*}(1-2\e^{-x_1^*})y_1 -dy_2,
  \end{split}
  \label{eq:8}
\end{equation}
where we considered $y_3 = 0$, since $y_3$ is any constant in the linearized flow, i.e., $\dot{y}_3=0$. Finally, considering the solutions of Eqs. (\ref{eq:8}) in the polar coordinates $y_1 = r\cos(\alpha)$ and $y_2 = r\sin(\alpha)$, we determine the angular frequency,
\begin{equation}
  \dot{\alpha} = \frac{1}{y_1^2+y_2^2}(y_1\dot{y_2}-y_2\dot{y_1}).
  \label{eq:9}
\end{equation}
Thus, by Eqs. (\ref{eq:2}) and (\ref{eq:3}), we can determine numerically the winding number inside a period-$m$ main-body window and then calculate the associated torsion number with Eq. (\ref{eq:4}), where $m$ is given by $m=T_\gamma/T$.

\begin{figure*}[!htp] \centering
\includegraphics[width=15cm,height=6cm]{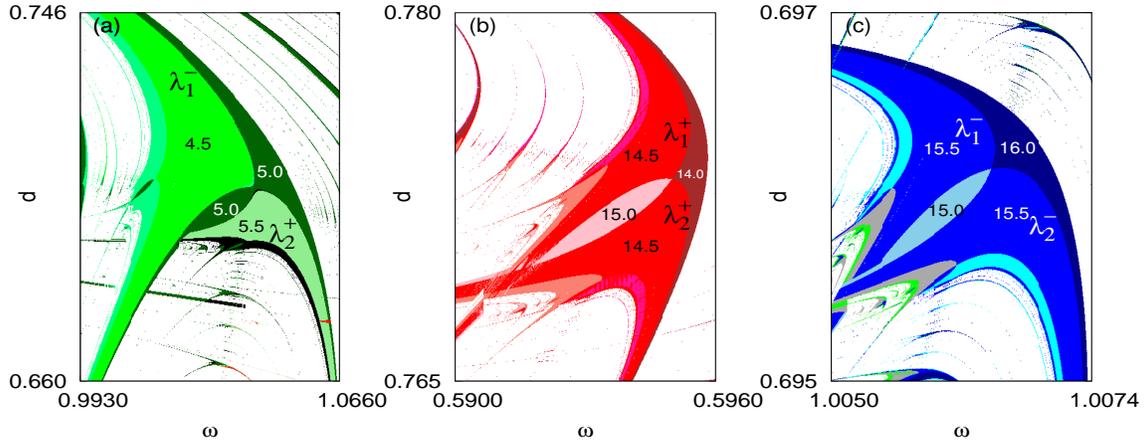}
\caption{(Color online) The white region indicates chaos while the colored regions indicate complex periodic windows according to the winding numbers of a grid of the system parameters. The numbers noted on the different winding number regions, $A$,$B$, $C$, and $D$, are the correspondent torsion numbers $n$. The $\lambda$ symbols indicate the window skeleton, and the $\lambda$ superscript $+$ or $-$ indicate increases or decreases in the torsion numbers.  (a) $n_A=n_B=5.0$, $n_C=5.5$, and $n_D=4.5$. (b) $n_C=n_D=14.5$, $n_A=14.0$, and $n_B=15.0$. (c) $n_C=n_D=15.5$, $n_A=16.0$, and $n_B=15.0$.}
\label{fig:3}
\end{figure*}

We recall that the orbit $\gamma$ is stable if the Floquet multipliers $\mu_i$
associated with the $mT$-shift map (or any multiple of $mT$) are
$|\mu_i|<1$. For $\mu_i \in \mathbb{R}$, if $\mu_i$ are positive, the
orbit $\gamma'$ converge to $\gamma$ with orientation preserving
(monotonically) and if $\mu_i$ are negative, $\gamma'$ converge to $\gamma$ with
orientation reversing (non monotonically) \cite{Kuznetsov2004}. Since the
multipliers are $|\mu_i|<1$ from the saddle-node bifurcation
$(\mu=1)$, in region $A$, to the period-doubling bifurcation
$(\mu=-1)$ in regions $C$ and $D$, the phase difference by
the interval time $\Delta t = mT$ computed between regions $A$ and $C$
(or $D$) is $\pi$ or, i.e., $|n_A-n_{C(D)}| = 1/2$, as can be seen in Fig. \ref{fig:3}.

\begin{figure*}[!htp] \centering
\includegraphics[width=12cm,height=8cm]{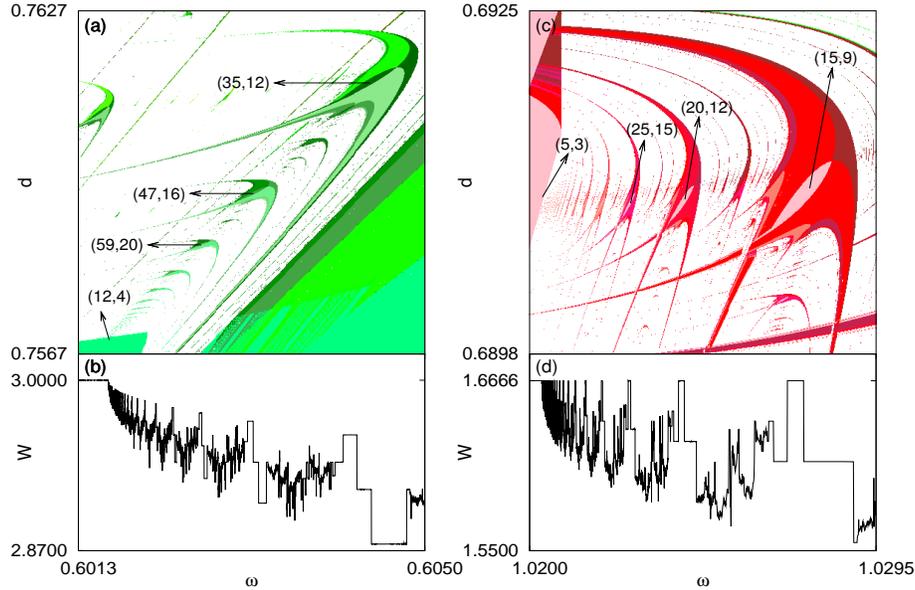}
\caption{(Color online) (a) Periodic window sequences obtained by computing the winding numbers. The pairs ($n_B, m_B$) noted on different periodic windows are the torsion number and the period of region $B$. The increments on the torsion number and periods are, respectively, $\tau=12.0$ and $\rho=4.0$. (b) Winding number, as a function of parameter $\omega$, calculated along a line crossing the $B$ regions of the window sequence of (a). Here,  the asymptotic winding number for this sequence is $w_\infty=3.0$, equal to its predicted value  $\tau/\rho=3.0$. (c) The same as (a) in another region of the parameter space (in this case $\tau=5.0$ and $\rho=3.0$). (d) The same as in (b) (in this case $w_\infty=\tau/\rho=1.6666$). The colors are the same meaning of Fig. \ref{fig:3}.}
\label{fig:4}
\end{figure*}

For the Morse oscillator, we obtain two-dimensional parameter spaces [shown in Figs. \ref{fig:3} and \ref{fig:4}] by computing the torsion and the winding numbers for each $d \times \omega$ parameters of a two-dimensional mesh of $500 \times 500$ equally spaced. We assign different colors to designate winding number values of periodic attractors, while the white color represents the parameters correspondent to chaotic attractors. With this procedure we identify
the uniform winding number regions, $A$, $B$, $C$, and $D$ indicated in Fig. \ref{fig:2}. We also identify inside
the three periodic windows, in Fig. \ref{fig:3}, the skeleton separating areas with uniform torsion numbers. To be more precise, lets define the left and right side of the curves $\lambda_1$ and $\lambda_2$ traversing the curve from the bottom to the top (see Fig. \ref{fig:2}). The torsion number is always increased or decreased by $1/2$ when we cross the curve $\lambda_1$ ($\lambda_2$) from the right (left) to the left (right). The superscripts $+$ and $-$, in
$\lambda_1$ and $\lambda_2$, indicate if the torsion number increases ($+$) or decreases ($-$) according to the defined
orientation. All the three possible skeleton are shown in Fig. \ref{fig:3}.

Figure 3 here.

Note that the main-body windows, shown in Fig. \ref{fig:3}, presents three different winding number values. In Fig. \ref{fig:3}(a), the torsion number decreases by $1/2$ from $A$ to $C$ ($\lambda_1 = \lambda_1^-$, according our convention) and increases by $1/2$ from $C$ to $B$ ($\lambda_2 = \lambda_2^+$). Thus, the torsion number difference between regions $A$ and $B$ is zero. The same is verified for the route $A\to D\to B$, then $n_{A} = n_{B}$. Since the parameter sets in this window correspond to orbits with the same period $m$, according Eq. (\ref{eq:5}), the winding number in regions $A$ and $B$ are the same. Similarly, we conclude that in Fig. \ref{fig:3}(b) and (c), where $n_{C} = n_{D}$, the winding number of regions $C$ and $D$ are the same. Therefore, for the $i$-th complex window (see Fig. \ref{fig:4}), the winding number is given by
\begin{equation}
  w_{R_i} = \frac{n_{A_1}+(i-1)\tau + (k-l)/2}{m_1+(i-1)\rho},
  \label{eq:10}
\end{equation}
where $k,l \in \{0,1,2\}$ are, respectively, the number of times that the curves type $\lambda^+$ and $\lambda^-$ are crossed from region $A$ to any region $R$ crossing $\lambda_1$ and $\lambda_2$ just one time. Equation (\ref{eq:10}) describes the internal regions in these periodic window sequence and obey the convergence limit established in Eq. (\ref{eq:6}). Moreover, Eq. (\ref{eq:10}) states that if one periodic windows presents $w_B = \tau/\rho$, then, the winding number of region $B$ in any periodic window in the sequence is $w_{B_i} = \tau/\rho$.

To illustrate the validity of our results, we display in Fig. \ref{fig:4}(a) sequences of periodic windows where all
windows have their internal regions separated by $\lambda_1^-$ and $\lambda_2^+$. For this sequence, according to Eq. (\ref{eq:10}) with $k=1$ and $l=1$, all periodic windows present different winding number values in their central region $B$. Additionally, we show in Fig. \ref{fig:4}(b) the winding number calculated along a line passing through the region $B$ of all windows that compose the sequence. It is clear that the winding numbers converge to $w_\infty = 3.0$. The torsion number and the period increment can also be determined, in Fig. \ref{fig:4}(a), by $\tau = n_{B_{i+1}}-n_{B_i} = 12$ and $\rho = m_{i+1}-m_i = 4$, respectively. Thus, the winding number convergence is in agreement with Eq. (\ref{eq:6}).

Figure 4 here.

In Fig. \ref{fig:4}(c), we show another sequence of periodic windows internally separated by $\lambda_1^+$ and $\lambda_2^+$. For this sequence of windows, we measure $\tau = 5$, $\rho = 3$, and the winding number $w_B = 5/3$. Thus, as predicted by Eqs. (\ref{eq:10}) with $k=2$ and $l=0$, we measure $w_{B_i} = 5/3$. From Fig. \ref{fig:4}(d) we also verify the limit $w_\infty= \tau/\rho = 5/3$. Figure \ref{fig:4} shows that the regions where the sequences are  accumulating have the same correspondent winding number $w_\infty$.

In conclusion, we report the existence of torsion-adding phenomenon in periodic window sequences of the driven nonlinear oscillators. Additionally, we formulate a general rule [Eq. (\ref{eq:10})] to obtain the winding number of any window belonging to the sequence. From this general rule we obtain the winding number asymptotic limit for any sequence ($w_\infty=\tau/\rho$), such ratio seems to be an universal property of dynamical systems, once it requires only the existence of the period and torsion adding phenomena. Moreover, since there are no general theory ensuring that period-adding sequences are composed of an infinite number of windows, this limit is a theoretical evidence of the existence of infinite windows in the sequences. 

Furthermore, we present numerical simulations for the Morse oscillator where the reported torsion-adding phenomenon and winding number asymptotic limit are verified. We also performed numerical analysis for other nonlinear oscillators described by Eq. (\ref{eq:1}), and verified that all results are in complete agreement with our theory.

The authors thank the following Brazilian government agencies for the partial financial support: FAPESP, CNPq, and Capes.


\begin{thebibliography}{27}%
\makeatletter
\providecommand \@ifxundefined [1]{%
 \@ifx{#1\undefined}
}%
\providecommand \@ifnum [1]{%
 \ifnum #1\expandafter \@firstoftwo
 \else \expandafter \@secondoftwo
 \fi
}%
\providecommand \@ifx [1]{%
 \ifx #1\expandafter \@firstoftwo
 \else \expandafter \@secondoftwo
 \fi
}%
\providecommand \natexlab [1]{#1}%
\providecommand \enquote  [1]{``#1''}%
\providecommand \bibnamefont  [1]{#1}%
\providecommand \bibfnamefont [1]{#1}%
\providecommand \citenamefont [1]{#1}%
\providecommand \href@noop [0]{\@secondoftwo}%
\providecommand \href [0]{\begingroup \@sanitize@url \@href}%
\providecommand \@href[1]{\@@startlink{#1}\@@href}%
\providecommand \@@href[1]{\endgroup#1\@@endlink}%
\providecommand \@sanitize@url [0]{\catcode `\\12\catcode `\$12\catcode
  `\&12\catcode `\#12\catcode `\^12\catcode `\_12\catcode `\%12\relax}%
\providecommand \@@startlink[1]{}%
\providecommand \@@endlink[0]{}%
\providecommand \url  [0]{\begingroup\@sanitize@url \@url }%
\providecommand \@url [1]{\endgroup\@href {#1}{\urlprefix }}%
\providecommand \urlprefix  [0]{URL }%
\providecommand \Eprint [0]{\href }%
\providecommand \doibase [0]{http://dx.doi.org/}%
\providecommand \selectlanguage [0]{\@gobble}%
\providecommand \bibinfo  [0]{\@secondoftwo}%
\providecommand \bibfield  [0]{\@secondoftwo}%
\providecommand \translation [1]{[#1]}%
\providecommand \BibitemOpen [0]{}%
\providecommand \bibitemStop [0]{}%
\providecommand \bibitemNoStop [0]{.\EOS\space}%
\providecommand \EOS [0]{\spacefactor3000\relax}%
\providecommand \BibitemShut  [1]{\csname bibitem#1\endcsname}%
\let\auto@bib@innerbib\@empty
%</preamble>
\bibitem [{\citenamefont {Kaneko}(1982)}]{Kaneko1982}%
  \BibitemOpen
  \bibfield  {author} {\bibinfo {author} {\bibfnamefont {K.}~\bibnamefont
  {Kaneko}},\ }\href@noop {} {\bibfield  {journal} {\bibinfo  {journal} {Prog.
  Theor. Phys.}\ }\textbf {\bibinfo {volume} {68}},\ \bibinfo {pages} {669}
  (\bibinfo {year} {1982})}\BibitemShut {NoStop}%
\bibitem [{\citenamefont {Ren}\ \emph {et~al.}(1997)\citenamefont {Ren},
  \citenamefont {Hu}, \citenamefont {Zhang}, \citenamefont {Wang},
  \citenamefont {Gong},\ and\ \citenamefont {Xu}}]{Ren1997}%
  \BibitemOpen
  \bibfield  {author} {\bibinfo {author} {\bibfnamefont {W.}~\bibnamefont
  {Ren}}, \bibinfo {author} {\bibfnamefont {S.~J.}\ \bibnamefont {Hu}},
  \bibinfo {author} {\bibfnamefont {B.~J.}\ \bibnamefont {Zhang}}, \bibinfo
  {author} {\bibfnamefont {F.~Z.}\ \bibnamefont {Wang}}, \bibinfo {author}
  {\bibfnamefont {Y.~F.}\ \bibnamefont {Gong}}, \ and\ \bibinfo {author}
  {\bibfnamefont {J.~X.}\ \bibnamefont {Xu}},\ }\href@noop {} {\bibfield
  {journal} {\bibinfo  {journal} {Intern. J. of Bifur. and Chaos,}\ }\textbf
  {\bibinfo {volume} {7}},\ \bibinfo {pages} {1867} (\bibinfo {year}
  {1997})}\BibitemShut {NoStop}%
\bibitem [{\citenamefont {Jia}\ \emph {et~al.}(2012)\citenamefont {Jia},
  \citenamefont {Gu}, \citenamefont {Li},\ and\ \citenamefont
  {Zhao}}]{Jia2012}%
  \BibitemOpen
  \bibfield  {author} {\bibinfo {author} {\bibfnamefont {B.}~\bibnamefont
  {Jia}}, \bibinfo {author} {\bibfnamefont {H.}~\bibnamefont {Gu}}, \bibinfo
  {author} {\bibfnamefont {L.}~\bibnamefont {Li}}, \ and\ \bibinfo {author}
  {\bibfnamefont {X.}~\bibnamefont {Zhao}},\ }\href@noop {} {\bibfield
  {journal} {\bibinfo  {journal} {Cogn. Neurodyn.}\ }\textbf {\bibinfo {volume}
  {6}},\ \bibinfo {pages} {89} (\bibinfo {year} {2012})}\BibitemShut {NoStop}%
\bibitem [{\citenamefont {Pei}\ \emph {et~al.}(1986)\citenamefont {Pei},
  \citenamefont {Guo}, \citenamefont {Wu},\ and\ \citenamefont
  {Chua}}]{Periodchua1986}%
  \BibitemOpen
  \bibfield  {author} {\bibinfo {author} {\bibfnamefont {L.~Q.}\ \bibnamefont
  {Pei}}, \bibinfo {author} {\bibfnamefont {F.}~\bibnamefont {Guo}}, \bibinfo
  {author} {\bibfnamefont {S.~X.}\ \bibnamefont {Wu}}, \ and\ \bibinfo {author}
  {\bibfnamefont {L.~O.}\ \bibnamefont {Chua}},\ }\href@noop {} {\bibfield
  {journal} {\bibinfo  {journal} {IEEE Trans. Circuits and Syst.}\ }\textbf
  {\bibinfo {volume} {CAS-33}},\ \bibinfo {pages} {438} (\bibinfo {year}
  {1986})}\BibitemShut {NoStop}%
\bibitem [{\citenamefont {Pereira}\ \emph {et~al.}(2012)\citenamefont
  {Pereira}, \citenamefont {Colli},\ and\ \citenamefont
  {Sartorelli}}]{bubblesarto2004}%
  \BibitemOpen
  \bibfield  {author} {\bibinfo {author} {\bibfnamefont {F.~A.~C.}\
  \bibnamefont {Pereira}}, \bibinfo {author} {\bibfnamefont {E.}~\bibnamefont
  {Colli}}, \ and\ \bibinfo {author} {\bibfnamefont {J.~C.}\ \bibnamefont
  {Sartorelli}},\ }\href@noop {} {\bibfield  {journal} {\bibinfo  {journal}
  {Chaos}\ }\textbf {\bibinfo {volume} {22}},\ \bibinfo {pages} {013135}
  (\bibinfo {year} {2012})}\BibitemShut {NoStop}%
\bibitem [{\citenamefont {Yasuda}\ and\ \citenamefont
  {Hoh}(1994)}]{Yasuda1994}%
  \BibitemOpen
  \bibfield  {author} {\bibinfo {author} {\bibfnamefont {Y.}~\bibnamefont
  {Yasuda}}\ and\ \bibinfo {author} {\bibfnamefont {K.}~\bibnamefont {Hoh}},\
  }\href@noop {} {\bibfield  {journal} {\bibinfo  {journal} {Electron. Commun.
  Japan}\ }\textbf {\bibinfo {volume} {77}},\ \bibinfo {pages} {654} (\bibinfo
  {year} {1994})}\BibitemShut {NoStop}%
\bibitem [{\citenamefont {Hauser}\ \emph {et~al.}(1997)\citenamefont {Hauser},
  \citenamefont {Olsen}, \citenamefont {Bronnikova},\ and\ \citenamefont
  {Schaffer}}]{Hauser1997}%
  \BibitemOpen
  \bibfield  {author} {\bibinfo {author} {\bibfnamefont {M.~J.~B.}\
  \bibnamefont {Hauser}}, \bibinfo {author} {\bibfnamefont {L.~F.}\
  \bibnamefont {Olsen}}, \bibinfo {author} {\bibfnamefont {T.~B.}\ \bibnamefont
  {Bronnikova}}, \ and\ \bibinfo {author} {\bibfnamefont {W.~M.}\ \bibnamefont
  {Schaffer}},\ }\href@noop {} {\bibfield  {journal} {\bibinfo  {journal} {J.
  Phys. Chem. B}\ }\textbf {\bibinfo {volume} {101}},\ \bibinfo {pages} {5075}
  (\bibinfo {year} {1997})}\BibitemShut {NoStop}%
\bibitem [{\citenamefont {Bonatto}\ and\ \citenamefont
  {Gallas}(2007)}]{Bonatto2007}%
  \BibitemOpen
  \bibfield  {author} {\bibinfo {author} {\bibfnamefont {C.}~\bibnamefont
  {Bonatto}}\ and\ \bibinfo {author} {\bibfnamefont {J.~A.~C.}\ \bibnamefont
  {Gallas}},\ }\href@noop {} {\bibfield  {journal} {\bibinfo  {journal} {Phys.
  Rev. E}\ }\textbf {\bibinfo {volume} {75}},\ \bibinfo {pages} {055204(R)}
  (\bibinfo {year} {2007})}\BibitemShut {NoStop}%
\bibitem [{\citenamefont {Albuquerque}\ and\ \citenamefont
  {Rech}(2012)}]{Albuquerque2012}%
  \BibitemOpen
  \bibfield  {author} {\bibinfo {author} {\bibfnamefont {H.~A.}\ \bibnamefont
  {Albuquerque}}\ and\ \bibinfo {author} {\bibfnamefont {P.~C.}\ \bibnamefont
  {Rech}},\ }\href@noop {} {\bibfield  {journal} {\bibinfo  {journal} {Int. J.
  Circ. Theor. Appl.}\ }\textbf {\bibinfo {volume} {40}},\ \bibinfo {pages}
  {189} (\bibinfo {year} {2012})}\BibitemShut {NoStop}%
\bibitem [{\citenamefont {Stegemann}\ \emph {et~al.}(2010)\citenamefont
  {Stegemann}, \citenamefont {Albuquerque},\ and\ \citenamefont
  {Rech}}]{Stegemann2010}%
  \BibitemOpen
  \bibfield  {author} {\bibinfo {author} {\bibfnamefont {C.}~\bibnamefont
  {Stegemann}}, \bibinfo {author} {\bibfnamefont {H.~A.}\ \bibnamefont
  {Albuquerque}}, \ and\ \bibinfo {author} {\bibfnamefont {P.~C.}\ \bibnamefont
  {Rech}},\ }\href@noop {} {\bibfield  {journal} {\bibinfo  {journal} {Chaos}\
  }\textbf {\bibinfo {volume} {20}},\ \bibinfo {pages} {023103} (\bibinfo
  {year} {2010})}\BibitemShut {NoStop}%
\bibitem [{\citenamefont {Gallas}(1993)}]{Gallas1993}%
  \BibitemOpen
  \bibfield  {author} {\bibinfo {author} {\bibfnamefont {J.~A.~C.}\
  \bibnamefont {Gallas}},\ }\href@noop {} {\bibfield  {journal} {\bibinfo
  {journal} {Phys. Rev. Lett.}\ }\textbf {\bibinfo {volume} {70}},\ \bibinfo
  {pages} {2714} (\bibinfo {year} {1993})}\BibitemShut {NoStop}%
\bibitem [{\citenamefont {Baptista}\ and\ \citenamefont
  {Caldas}(1996)}]{Murilo1996}%
  \BibitemOpen
  \bibfield  {author} {\bibinfo {author} {\bibfnamefont {M.~S.}\ \bibnamefont
  {Baptista}}\ and\ \bibinfo {author} {\bibfnamefont {I.~L.}\ \bibnamefont
  {Caldas}},\ }\href@noop {} {\bibfield  {journal} {\bibinfo  {journal} {Chaos,
  Solitons and Fractals}\ }\textbf {\bibinfo {volume} {7}},\ \bibinfo {pages}
  {325} (\bibinfo {year} {1996})}\BibitemShut {NoStop}%
\bibitem [{\citenamefont {Stoop}\ \emph {et~al.}(2010)\citenamefont {Stoop},
  \citenamefont {Benner},\ and\ \citenamefont {Uwate}}]{Stoop2010}%
  \BibitemOpen
  \bibfield  {author} {\bibinfo {author} {\bibfnamefont {R.}~\bibnamefont
  {Stoop}}, \bibinfo {author} {\bibfnamefont {P.}~\bibnamefont {Benner}}, \
  and\ \bibinfo {author} {\bibfnamefont {Y.}~\bibnamefont {Uwate}},\
  }\href@noop {} {\bibfield  {journal} {\bibinfo  {journal} {Phys. Rev. Lett.}\
  }\textbf {\bibinfo {volume} {105}},\ \bibinfo {pages} {074102} (\bibinfo
  {year} {2010})}\BibitemShut {NoStop}%
\bibitem [{\citenamefont {Parlitz}\ and\ \citenamefont
  {Lauterborn}(1985{\natexlab{a}})}]{Parlitz1985a}%
  \BibitemOpen
  \bibfield  {author} {\bibinfo {author} {\bibfnamefont {U.}~\bibnamefont
  {Parlitz}}\ and\ \bibinfo {author} {\bibfnamefont {W.}~\bibnamefont
  {Lauterborn}},\ }\href@noop {} {\bibfield  {journal} {\bibinfo  {journal} {Z.
  Naturforsch}\ }\textbf {\bibinfo {volume} {41}},\ \bibinfo {pages} {605}
  (\bibinfo {year} {1985}{\natexlab{a}})}\BibitemShut {NoStop}%
\bibitem [{\citenamefont {Parlitz}\ and\ \citenamefont
  {Lauterborn}(1985{\natexlab{b}})}]{Parlitz1985b}%
  \BibitemOpen
  \bibfield  {author} {\bibinfo {author} {\bibfnamefont {U.}~\bibnamefont
  {Parlitz}}\ and\ \bibinfo {author} {\bibfnamefont {W.}~\bibnamefont
  {Lauterborn}},\ }\href@noop {} {\bibfield  {journal} {\bibinfo  {journal}
  {Phys. Lett.}\ }\textbf {\bibinfo {volume} {107A}},\ \bibinfo {pages} {351}
  (\bibinfo {year} {1985}{\natexlab{b}})}\BibitemShut {NoStop}%
\bibitem [{\citenamefont {Parlitz}\ \emph {et~al.}(1990)\citenamefont
  {Parlitz}, \citenamefont {Englisch}, \citenamefont {Scheffczyk},\ and\
  \citenamefont {Lauterborn}}]{Parlitz1990}%
  \BibitemOpen
  \bibfield  {author} {\bibinfo {author} {\bibfnamefont {U.}~\bibnamefont
  {Parlitz}}, \bibinfo {author} {\bibfnamefont {V.}~\bibnamefont {Englisch}},
  \bibinfo {author} {\bibfnamefont {C.}~\bibnamefont {Scheffczyk}}, \ and\
  \bibinfo {author} {\bibfnamefont {W.}~\bibnamefont {Lauterborn}},\
  }\href@noop {} {\bibfield  {journal} {\bibinfo  {journal} {J. Acoust. Soc.
  Am.}\ }\textbf {\bibinfo {volume} {88}},\ \bibinfo {pages} {1061} (\bibinfo
  {year} {1990})}\BibitemShut {NoStop}%
\bibitem [{\citenamefont {Englisch}\ and\ \citenamefont
  {Lauterborn}(1991)}]{Englisch1991}%
  \BibitemOpen
  \bibfield  {author} {\bibinfo {author} {\bibfnamefont {V.}~\bibnamefont
  {Englisch}}\ and\ \bibinfo {author} {\bibfnamefont {W.}~\bibnamefont
  {Lauterborn}},\ }\href@noop {} {\bibfield  {journal} {\bibinfo  {journal}
  {Phys. Rev. A}\ }\textbf {\bibinfo {volume} {44}},\ \bibinfo {pages} {916}
  (\bibinfo {year} {1991})}\BibitemShut {NoStop}%
\bibitem [{\citenamefont {Parlitz}(1993)}]{Parlitz1993}%
  \BibitemOpen
  \bibfield  {author} {\bibinfo {author} {\bibfnamefont {U.}~\bibnamefont
  {Parlitz}},\ }\href@noop {} {\bibfield  {journal} {\bibinfo  {journal} {Int.
  J. Of Bifurc. and Chaos}\ }\textbf {\bibinfo {volume} {3}},\ \bibinfo {pages}
  {703} (\bibinfo {year} {1993})}\BibitemShut {NoStop}%
\bibitem [{Not()}]{Nota1}%
  \BibitemOpen
  \href@noop {} {}\bibinfo {note} {The stroboscopic Poincar\'e map defined by
  the driven oscillator period $T=2\pi/\omega$}\BibitemShut {NoStop}%
\bibitem [{\citenamefont {Glass}\ \emph {et~al.}(1984)\citenamefont {Glass},
  \citenamefont {Guevara}, \citenamefont {Belair},\ and\ \citenamefont
  {Shrier}}]{Glass1984}%
  \BibitemOpen
  \bibfield  {author} {\bibinfo {author} {\bibfnamefont {L.}~\bibnamefont
  {Glass}}, \bibinfo {author} {\bibfnamefont {M.~R.}\ \bibnamefont {Guevara}},
  \bibinfo {author} {\bibfnamefont {J.}~\bibnamefont {Belair}}, \ and\ \bibinfo
  {author} {\bibfnamefont {A.}~\bibnamefont {Shrier}},\ }\href@noop {}
  {\bibfield  {journal} {\bibinfo  {journal} {Phys. Rev. A}\ }\textbf {\bibinfo
  {volume} {29}},\ \bibinfo {pages} {1348} (\bibinfo {year}
  {1984})}\BibitemShut {NoStop}%
\bibitem [{\citenamefont {Belair}\ and\ \citenamefont
  {Glass}(1985)}]{Belair1985}%
  \BibitemOpen
  \bibfield  {author} {\bibinfo {author} {\bibfnamefont {J.}~\bibnamefont
  {Belair}}\ and\ \bibinfo {author} {\bibfnamefont {L.}~\bibnamefont {Glass}},\
  }\href@noop {} {\bibfield  {journal} {\bibinfo  {journal} {Physica D}\
  }\textbf {\bibinfo {volume} {16}},\ \bibinfo {pages} {143} (\bibinfo {year}
  {1985})}\BibitemShut {NoStop}%
\bibitem [{\citenamefont {Kuznetsov}(2004)}]{Kuznetsov2004}%
  \BibitemOpen
  \bibfield  {author} {\bibinfo {author} {\bibfnamefont {Y.~A.}\ \bibnamefont
  {Kuznetsov}},\ }\href@noop {} {\emph {\bibinfo {title} {Elements of applied
  bifurcation theory}}}\ (\bibinfo  {publisher} {Springer},\ \bibinfo {address}
  {United States of America},\ \bibinfo {year} {2004})\BibitemShut {NoStop}%
\bibitem [{\citenamefont {Bonatto}\ \emph {et~al.}(2008)\citenamefont
  {Bonatto}, \citenamefont {Gallas},\ and\ \citenamefont
  {Ueda}}]{Bonatto2008a}%
  \BibitemOpen
  \bibfield  {author} {\bibinfo {author} {\bibfnamefont {C.}~\bibnamefont
  {Bonatto}}, \bibinfo {author} {\bibfnamefont {J.~A.~C.}\ \bibnamefont
  {Gallas}}, \ and\ \bibinfo {author} {\bibfnamefont {Y.}~\bibnamefont
  {Ueda}},\ }\href@noop {} {\bibfield  {journal} {\bibinfo  {journal} {Phys.
  Rev. E}\ }\textbf {\bibinfo {volume} {77}},\ \bibinfo {pages} {026217}
  (\bibinfo {year} {2008})}\BibitemShut {NoStop}%
\bibitem [{\citenamefont {Bonatto}\ and\ \citenamefont
  {Gallas}(2008)}]{Bonatto2008b}%
  \BibitemOpen
  \bibfield  {author} {\bibinfo {author} {\bibfnamefont {C.}~\bibnamefont
  {Bonatto}}\ and\ \bibinfo {author} {\bibfnamefont {J.~A.~C.}\ \bibnamefont
  {Gallas}},\ }\href@noop {} {\bibfield  {journal} {\bibinfo  {journal} {Phil.
  Trans. R. Soc. A}\ }\textbf {\bibinfo {volume} {366}},\ \bibinfo {pages}
  {505} (\bibinfo {year} {2008})}\BibitemShut {NoStop}%
\bibitem [{\citenamefont {de~Souza}\ \emph {et~al.}(2012)\citenamefont
  {de~Souza}, \citenamefont {Lima}, \citenamefont {Caldas}, \citenamefont
  {Medrano-T.},\ and\ \citenamefont {Guimarães-Filho}}]{Souza2012}%
  \BibitemOpen
  \bibfield  {author} {\bibinfo {author} {\bibfnamefont {S.~L.~T.}\
  \bibnamefont {de~Souza}}, \bibinfo {author} {\bibfnamefont {A.~A.}\
  \bibnamefont {Lima}}, \bibinfo {author} {\bibfnamefont {I.~L.}\ \bibnamefont
  {Caldas}}, \bibinfo {author} {\bibfnamefont {R.~O.}\ \bibnamefont
  {Medrano-T.}}, \ and\ \bibinfo {author} {\bibfnamefont {Z.~O.}\ \bibnamefont
  {Guimarães-Filho}},\ }\href@noop {} {\bibfield  {journal} {\bibinfo
  {journal} {Phys. Lett. A}\ }\textbf {\bibinfo {volume} {376}},\ \bibinfo
  {pages} {1290} (\bibinfo {year} {2012})}\BibitemShut {NoStop}%
\bibitem [{\citenamefont {Xian-Feng}\ and\ \citenamefont
  {A.~Y.-T.~Leung}(2012)}]{Feng2012}%
  \BibitemOpen
  \bibfield  {author} {\bibinfo {author} {\bibfnamefont {L.}~\bibnamefont
  {Xian-Feng}}\ and\ \bibinfo {author} {\bibfnamefont {C.~Y.-D.}\ \bibnamefont
  {A.~Y.-T.~Leung}},\ }\href@noop {} {\bibfield  {journal} {\bibinfo  {journal}
  {Chin. Phys. Lett.}\ }\textbf {\bibinfo {volume} {29}},\ \bibinfo {pages}
  {010201} (\bibinfo {year} {2012})}\BibitemShut {NoStop}%
\bibitem [{\citenamefont {Scheffczyk}\ \emph {et~al.}(1991)\citenamefont
  {Scheffczyk}, \citenamefont {Parlitz}, \citenamefont {Kurz}, \citenamefont
  {Konp},\ and\ \citenamefont {Lauterborn}}]{Scheffczyk1991}%
  \BibitemOpen
  \bibfield  {author} {\bibinfo {author} {\bibfnamefont {C.}~\bibnamefont
  {Scheffczyk}}, \bibinfo {author} {\bibfnamefont {U.}~\bibnamefont {Parlitz}},
  \bibinfo {author} {\bibfnamefont {T.}~\bibnamefont {Kurz}}, \bibinfo {author}
  {\bibfnamefont {W.}~\bibnamefont {Konp}}, \ and\ \bibinfo {author}
  {\bibfnamefont {W.}~\bibnamefont {Lauterborn}},\ }\href@noop {} {\bibfield
  {journal} {\bibinfo  {journal} {Phys. Rev. A}\ }\textbf {\bibinfo {volume}
  {43}},\ \bibinfo {pages} {6495} (\bibinfo {year} {1991})}\BibitemShut
  {NoStop}%
\end{thebibliography}
\end{document}